\documentclass[aps,prl,twocolumn,groupedaddress,showpacs]{revtex4}
\usepackage{graphicx}
\graphicspath{{figure file fold path}}
\usepackage[dvipsnames,usenames]{color}
\usepackage{color}
\usepackage{amsmath}
\usepackage{float}
\usepackage{amssymb}
\usepackage{amsthm}
\usepackage{mathrsfs}
\usepackage{tabularray}
\usepackage{array}
\usepackage{booktabs}
\usepackage{siunitx}

\emergencystretch=\maxdimen
\begin{document}
\title{Inverse Design of Tunable Infrared Metasurface Absorbers via a Conditional Wasserstein Generative Adversarial Network}
\author{Hui Shen$^{1}$, Tong Wang$^{2}$, Xiaoqiang Yao$^{2}$, Ouling Wu$^{3}$, Changjian Xie$^{1}$, Chao Qian$^{3}$, Hongsheng Chen$^{3}$, and Tao Wang$^{1}$}

\affiliation{$^1$State Key Laboratory of Integrated Service Networks, Xidian University, Xi'an 710071, China}
\affiliation{$^2$Air Defense and Antimissile School, Air Force Engineering University, Xi’an 710051, China}
\affiliation{$^3$Inter disciplinary Center for Quantum Information, College of Information Science and Electronic Engineering, Zhejiang University, Hangzhou 310027, China}

\date{\today}

\begin{abstract}
Narrowband perfect absorbers are interesting for spectrum sensing, molecular detection, and infrared imaging. However, their design remains constrained by intuitive, iterative methods that lack flexibility, while also facing challenges in multi-objective optimization. Here, we introduce a deep learning-enabled inverse-design framework that overcomes these limitations through a conditional Wasserstein Generative Adversarial Network (WGAN). The main contribution of this work is a dual-channel image encoding scheme that jointly represents the geometry and thickness of a Si$_3$N$_4$ meta-layer, facilitating the network to learn the distribution of viable structures for a target optical response. This approach naturally solves the inherent ``one-to-many'' design issue, giving a diverse portfolio of functional candidates from a single input spectrum. The designed absorbers achieve exceptional spectral fidelity, with resonance peak errors below 5 nm, a mean squared error (MSE) on the order of $10^{-3}$, and the capacity to produce over 10 distinct, high-performance designs per target. Furthermore, we demonstrate the model's robustness under oblique illumination, showing that it can be efficiently fine-tuned to maintain spectral accuracy across incidence angles from $10^\circ$ to $40^\circ$ by transfer learning, thus extending its practical utility to non-normal operating conditions. Full-wave simulations confirm that the generated geometries support a hybrid plasmonic-dielectric resonance, leading to near-perfect absorption and strong near-field enhancement. Our study provides a robust, physics-aware design paradigm that moves beyond conventional parametric optimization. The introduced framework establishes a versatile platform for the on-demand inverse design of advanced photonic devices for sensing, spectroscopy, and optical signal processing.
\end{abstract}


\maketitle 

\section{Introduction and objectives}
Metasurface absorbers have emerged as versatile platforms for manipulating electromagnetic waves at subwavelength scales~\cite{meng2021optical, zeng2025performance}. By tailoring the geometry, spatial arrangement, and material composition of meta-atoms, the structures can achieve impedance matching with free space, resulting in ultra-high absorption across various spectral ranges~\cite{butt2025ultra}. While significant progresses have been made in broadband absorption designs, narrowband metasurface absorbers offer distinct advantages for applications such as spectral sensing, filtering, and molecular fingerprinting~\cite{huang2023ultrasensitive, cui2024roadmap, rakhshani2025metamaterial}. Recent studies have demonstrated ultra-narrowband responses using geometric phase resonance or wavelength-selective meta-atoms~\cite{ouyang2025ultra, jung2015wavelength}. Nevertheless, such designs often lack post-fabrication tunability and require labor-intensive parameter sweeps to adjust resonance peaks, usually at the cost of absorption efficiency or bandwidth.

The design of high-performance photonic devices involves exploring a vast, high-dimensional parameter space~\cite{Melati2025, saifullah2025deep}. Conventional design strategies heavily rely on empirical intuition and local optimization, and thus rarely deliver globally optimal solutions. Furthermore, typically restricted to narrow regions of the design landscape, such methods often necessitate labor-intensive, computationally costly parameter sweeps, yet may still overlook unconventional but high-performing configurations~\cite{so2019designing, an2021multifunctional}.

Inverse design methodologies~\cite{wiecha2021deep, li2022empowering}, particularly those driven by deep learning, have emerged as powerful alternatives that reframe the design process by directly mapping desired optical responses to feasible structural parameters~\cite{liu2021tackling, wiecha2021deep, qian2025guidance}. By learning complex, nonlinear relationships from large datasets, data-driven models can efficiently explore the design space beyond the traditional simulation limits. Among these, generative adversarial networks (GANs) have proven highly effective in capturing intricate distributions of photonic geometries and generating diverse, high-performance candidate structures that satisfy target optical behaviors~\cite{dash2023review, alqahtani2021applications}. Notably, conditional and Wasserstein GAN variants have enhanced training stability and output quality, enabling robust solutions to inherently ill-posed ``one-to-many'' design issues where multiple structures can produce similar spectral responses~\cite{an2021multifunctional}.

In this work, we introduce a WGAN-based inverse design framework for designing narrowband infrared metasurface absorbers with customizable resonance peaks. Our method encodes structural geometry and dielectric thickness into a dual-channel image format, supporting efficient learning of the structure-spectrum relationship. The trained model not only generates high-fidelity designs, but also naturally resolves the one-to-many mapping challenge, providing multiple feasible structures for a single target spectrum. The generated designs are validated through finite-difference time-domain (FDTD) simulations, confirming both spectral accuracy and physical plausibility. Furthermore, we demonstrate the model's adaptability to oblique incidence angles via efficient fine-tuning, highlighting its robustness and practical utility in real-world illumination scenarios.

\section{Methodology}
\subsection{Metasurface Architecture and Dataset Preparation}
\begin{figure}[ht!]
\centering
  \includegraphics[width=8.3cm]{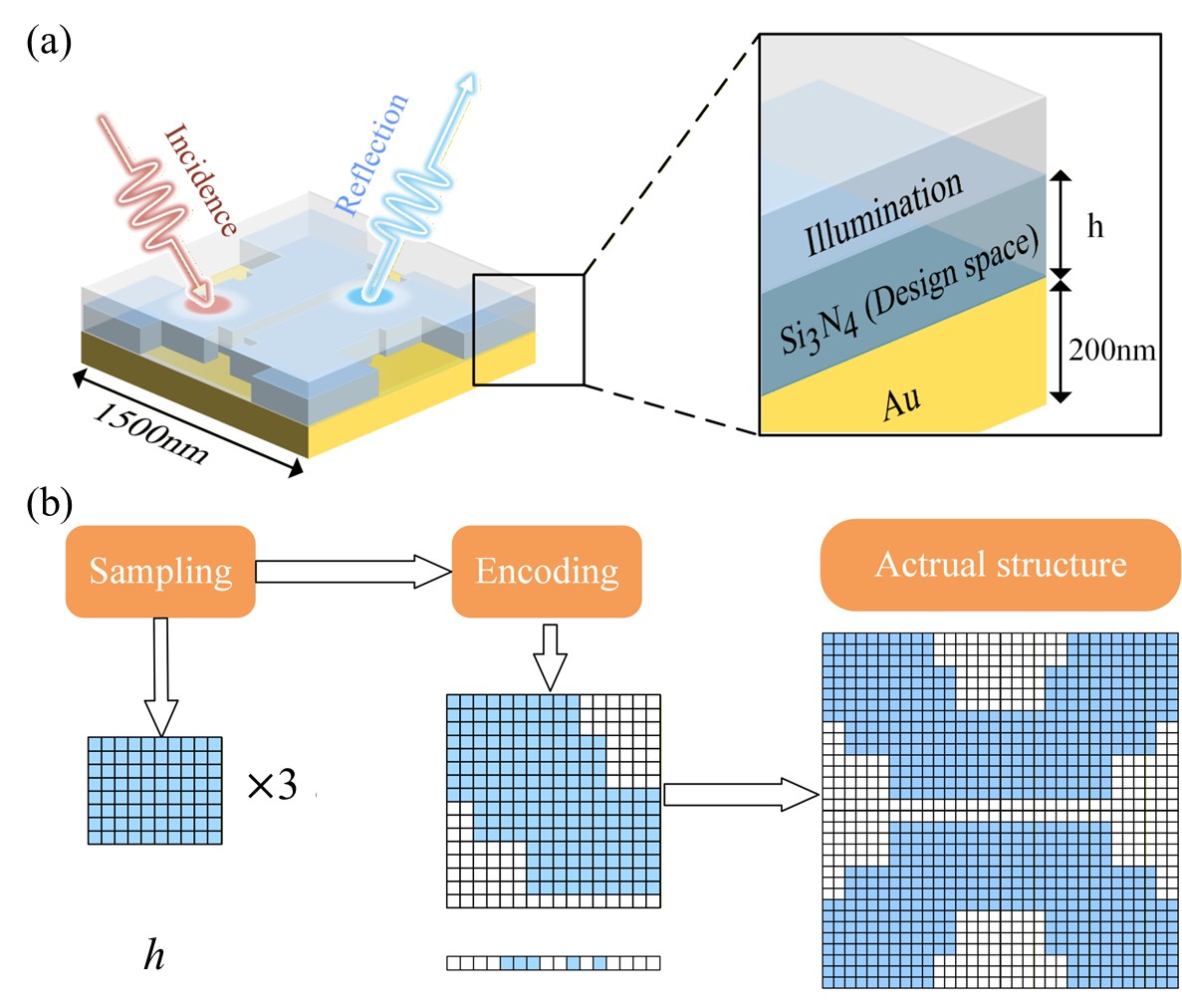}
  \caption{Schematic diagram of the proposed structure (a) and sampling algorithms and encoding methods for geometric shapes (b).}
  \label{Structure_and_encoding}
\end{figure}

The metasurface absorber, illustrated in Fig.~\ref{Structure_and_encoding}a, is constructed as a layered stack on a silica (SiO$_2$) substrate. A gold (Au) thin film with 200 nm-thickness serves as a bottom reflector, above which resides a patterned silicon nitride (Si$_3$N$_4$) dielectric layer. The unit cell period is set to 1.5~$\mu$m. Within this configuration, the geometry and thickness of the Si$_3$N$_4$ layer constitute the tunable design variables for inverse optimization, while the gold film acts as an optically opaque back mirror, effectively suppressing transmission and confining incident light to facilitate strong resonant absorption. To parameterize the design space, each unit cell is discretized into a $32 \times 32$ grid, with a pixel dimension of approximately 50 nm (Fig.~\ref{Structure_and_encoding}b), which is well within the resolution capabilities of state-of-the-art nanofabrication techniques~\cite{sortino2025atomic}.

To reduce the complexity of the design, only axisymmetric structures are considered, meaning the relevant geometric information is contained in one quadrant of the full grid. Next, geometric shapes are created via a stochastic stacking process that mimics droplet coalescence. In particular, three rectangular sampling units of size $10 \times 8$ are defined, each with randomly assigned vertex coordinates. These units are superimposed to form continuous, naturally varying shapes, which are then binarized into arrays to serve as the geometry training set. This approach maintains both global shape continuity and fine local details, promoting structural diversity and physical plausibility.

In parallel, the thickness of the Si$_3$N$_4$ layer is encoded as a binary vector and placed centrally within a separate $16 \times 16$ channel to mitigate convolutional edge effects~\cite{ferraz2023benchmarking}. The geometric layout and thickness channels are concatenated into a dual-channel image, which, along with the target absorption spectrum, forms a complete training instance. A total of 32,000 such instances are generated, partitioned 9:1 into training and testing subsets to support robust model development and evaluation.

While representing the Si$_3$N$_4$ layer thickness as a single value might intuitively seem to simplify network training, the adopted channel-encoding approach offers several advantages over direct scalar inputs: First, due to the nanoscale dimensions involved, the numerical differences between decimal thickness values are relatively small. In contrast, geometric variations are encoded as differences in a $16 \times 16$ binary vector, where the magnitude of individual element differences is on the order of 1. This disparity in input scales can lead to an imbalance in how the network weights geometric and thickness information during initial training, potentially prolonging convergence. Second, encoding both thickness and geometry into image channels allows the data to be treated uniformly as spatial information. This formulation requires only minimal modification to the input channels of established image-based models, enabling the direct use of state-of-the-art image processing architectures and training techniques without substantially increasing computational overhead. Third, distributing the decimal thickness information across multiple binary pixels provides a more robust and structured representation. This helps mitigate noise-related artifacts often encountered in GANs by supplying the discriminator with richer, more discriminative features to distinguish between realistic and generated structural images~\cite{jiang2019global, so2019designing}. 

\subsection{Conditional Wasserstein GAN Architecture}
The inverse design framework is built upon a conditional Wasserstein Generative Adversarial Network (WGAN), a deep learning architecture designed to learn and replicate the complex, high-dimensional distribution of viable metasurface geometries corresponding to desired optical responses~\cite{liu2018training, jiang2021deep}. This approach fundamentally reframes the design problem from a deterministic search to a probabilistic generative process, enabling efficient exploration of the design space.

Standard GANs, while powerful for generative tasks, are notoriously difficult to train due to issues like mode collapse and vanishing gradients, often leading to unstable convergence and poor sample diversity~\cite{goodfellow2014generative}. The WGAN variant mitigates these problems by employing the Wasserstein distance (Earth-Mover distance) as its training objective instead of the Jensen-Shannon divergence. This metric provides a smoother and more meaningful gradient signal across the entire training process, even when the distributions of real and generated data have minimal overlap~\cite{salimans2016improved}. As illustrated in Fig.~\ref{WGAN_and_algrithm}a, our conditional WGAN comprises two neural networks engaged in an adversarial minimax game: a \textit{Generator} (G) and a \textit{Discriminator} (D), also referred to as a critic in the WGAN framework~\cite{cretu2024synthesis}.

\subsubsection*{Generator Architecture}
The Generator (G) acts as the design synthesis engine. Its purpose is to map a conditional input, comprising a target absorption spectrum and a random latent vector, to a realistic, dual-channel structural image representing the Si$_3$N$_4$ geometry and thickness~\cite{roy2023novel}. The input is a concatenated vector of (1) the target absorption spectrum $y$ (a 1D array over the operational wavelength band), and (2) a noise vector $z$ sampled from a standard normal distribution $\mathcal{N}(0, I)$. The noise provides the stochasticity necessary for generating diverse structures for a single spectral target~\cite{wang2021spectral, sridharan2022deep}.

The network is based on a deep convolutional decoder. The combined input, which undergoes dimension alignment via replication, is passed through a sequence of four transposed convolutional layers. Each layer progressively upsamples the feature maps, transforming the low-dimensional latent code into a high-resolution $16 \times 16$ spatial representation. Each transposed convolutional layer is followed by batch normalization to stabilize and accelerate training. LeakyReLU activations ($\alpha = 0.2$) are used in the hidden layers to prevent gradient vanishing. The final output layer uses a Tanh activation function to constrain pixel values to the range $[-1, 1]$, which are subsequently threshold to produce the final binary design mask and the continuous thickness-encoded channel.

\subsubsection*{Discriminator (Critic) Architecture}
The Discriminator (D) functions as a trainable, differentiable critic that learns to score the ``realism'' of a given structure. Unlike a traditional classifier, it outputs a scalar score rather than a probability. The discriminator takes a paired input: the dual-channel structural image and the corresponding target spectrum $y$. The spectrum is tiled and appended as additional channels to condition the critic's evaluation, ensuring the generated structure is judged in the context of its intended optical function~\cite{zeng2025anchor}.

The discriminator is a convolutional neural network comprising four strided convolutional layers~\cite{isola2017image}. These layers successively downsample the spatial dimensions while extracting hierarchical features, from local geometric details to global structural patterns~\cite{gulrajani2017improved}. Instance normalization is used instead of batch normalization in the critic, as recommended for WGAN training stability. LeakyReLU activations are employed throughout. The final layer is a convolutional layer that outputs a single scalar critic value $D(x, y)$. A higher score indicates the critic judges the input pair $(x, y)$ to be more likely from the true data distribution.

\subsubsection*{Loss Functions and Training}
The training objective is defined by the Wasserstein distance. The loss functions for the generator and critic are:

\begin{eqnarray}
\mathcal{L}_D &=& \mathbb{E}_{\tilde{x} \sim \mathbb{P}_{g}}\left[ D(\tilde{x}, y) \right] - \mathbb{E}_{x \sim \mathbb{P}_{r}}\left[ D(x, y) \right] \,\\
\mathcal{L}_G &=& -\mathbb{E}_{\tilde{x} \sim \mathbb{P}_{g}}\left[ D(\tilde{x}, y) \right]
\end{eqnarray}

\noindent where $\mathbb{P}_r$ is the distribution of real (simulated) structure-spectrum pairs, $\mathbb{P}_g$ is the generator's distribution, $x$ is a real sample, and $\tilde{x} = G(z, y)$ is a generated sample.

\begin{figure}[ht!]
\centering
  \includegraphics[width=8.5cm]{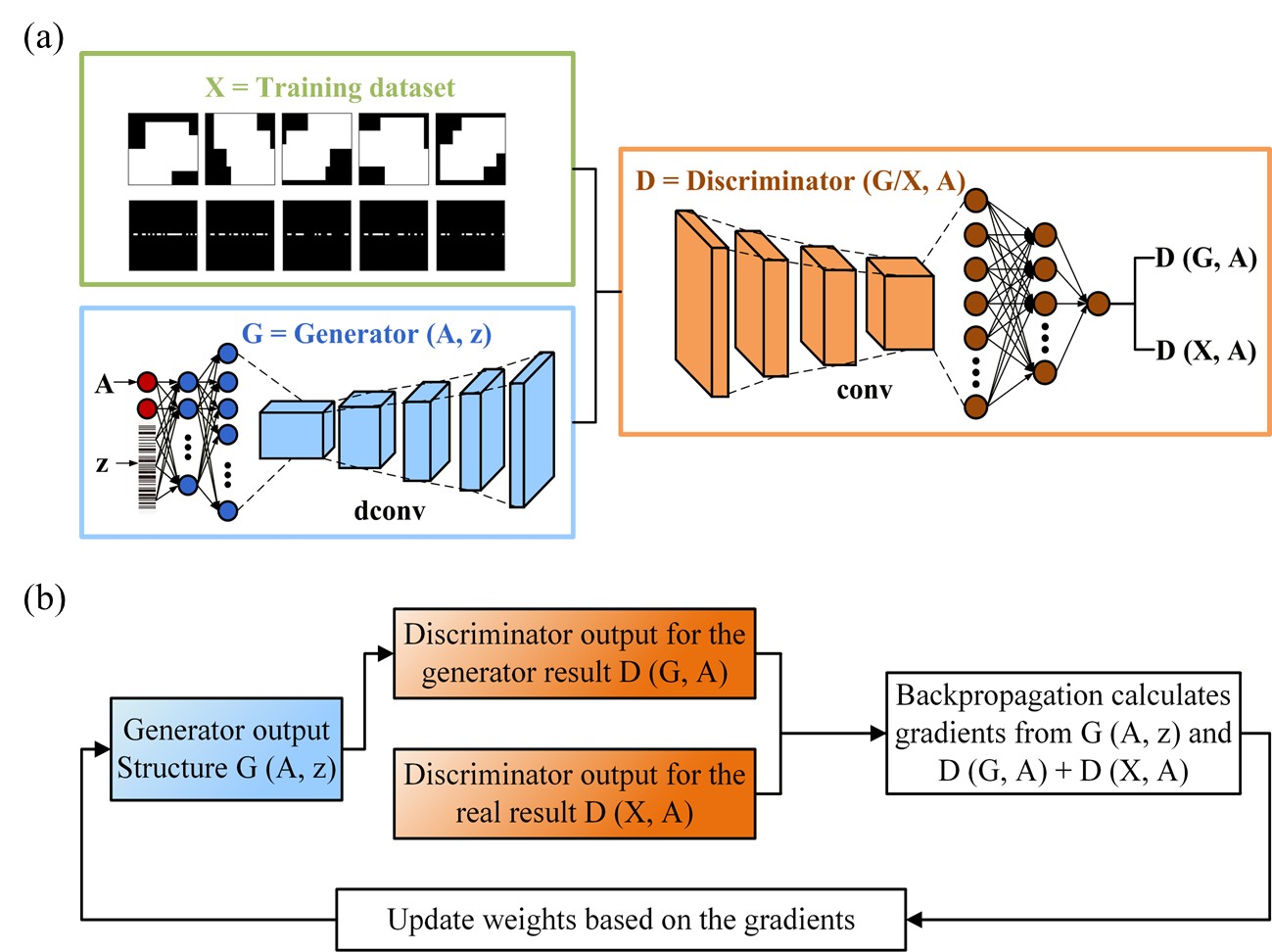}
  \caption{Architecture and training schematic of the conditional Wasserstein Generative Adversarial Network (WGAN). (a) The WGAN framework consists of a generator $G$ and a discriminator (critic) $D$. (b) The training algorithm illustrates the adversarial loop: the generator produces a candidate structure $G(z, y)$; the discriminator computes scores for both the generated output $D(G(z, y), y)$ and a real sample $D(x, y)$; gradients are calculated via backpropagation from the combined critic loss; and the weights of both networks are updated alternately to minimize the Wasserstein distance between the real and generated distributions. This process enables stable and diverse inverse design of metasurface absorbers.
}
  \label{WGAN_and_algrithm}
\end{figure}

To enforce the Lipschitz constraint required for a valid Wasserstein distance calculation, a gradient penalty term is added to the critic's loss. The training algorithm, depicted in Fig.~\ref{WGAN_and_algrithm}b, follows an adversarial loop: (1) The critic is updated multiple times (e.g., 5 times) per generator update to ensure it remains near optimal; (2) For each update, it evaluates both a batch of real pairs $(x, y)$ and generated pairs $(G(z, y), y)$, calculating $\mathcal{L}_D$; (3) The generator is then updated to minimize $\mathcal{L}_G$, which is equivalent to maximizing the critic's score for its generated structures, thereby ``fooling'' the critic.

This iterative process pushes $G$ to produce structures that are not only spectrally accurate but also indistinguishable from real designs according to the learned data distribution, naturally solving the ``one-to-many'' inverse problem inherent to photonics.

\section{Results and analysis}
\subsection{Spectral Accuracy and Design Fidelity}
The trained WGAN was tested on four target absorption peaks centered at 1440, 1480, 1520, and 1560 nm. For each target, ten noise vectors were sampled to generate candidate structures, which were then simulated using FDTD. The design with the lowest MSE between simulated and target spectra was selected.

\begin{figure*}[ht!]
\centering
  \includegraphics[width=12cm]{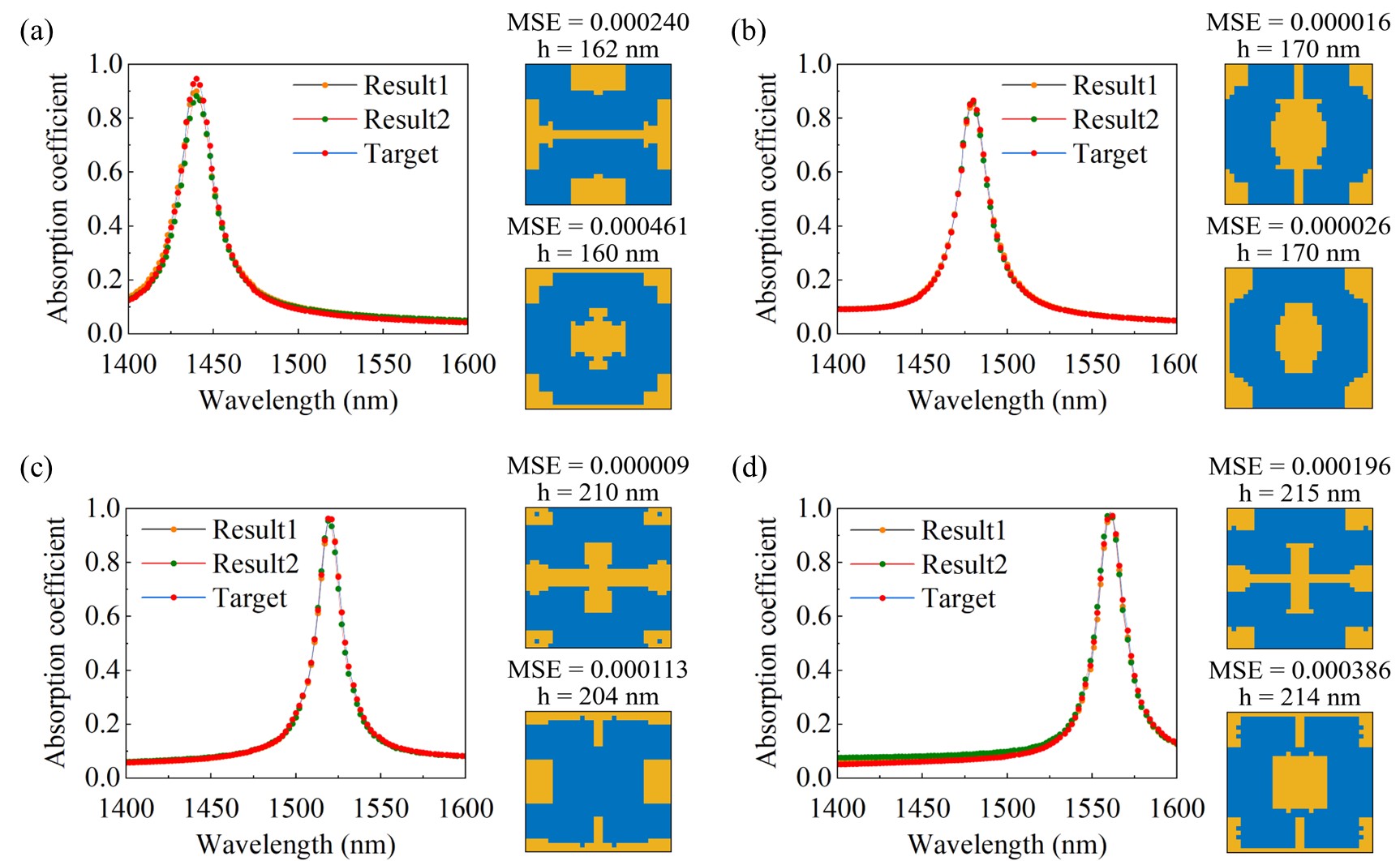}
  \caption{Inverse design results and spectral validation for target absorption peaks at (a) 1440, (b) 1480, (c) 1520, and (d) 1560 nm. For each target resonance (red dashed curve), two independently generated metasurface designs are shown alongside their corresponding numerically simulated absorption spectra (orange and green dash curves).
}
  \label{Absorbtion_spectrum}
\end{figure*}

Fig.~\ref{Absorbtion_spectrum} presents the performance evaluation of the WGAN model on inverse design for four distinct target absorption peaks. In specific, it shows a comprehensive analysis of the model's capability to generate functional metasurface absorbers targeting resonant wavelengths at 1440 nm, 1480 nm, 1520 nm, and 1560 nm. For each target (red dashed curves), two independently generated structural designs are shown alongside their corresponding full-wave simulated absorption spectra (orange and green dash curves). The predicted spectra demonstrate excellent agreement with the target resonances, with peak position errors consistently below 5 nm and mean squared error (MSE) values on the order of $10^{-4}$ to $10^{-3}$, validating the high spectral fidelity of the inverse design framework.

The paired structural solutions for each target, visually distinct in both geometric pattern and silicon nitride thickness (decoded and annotated above each design), indicate the model's successful learning of the non-unique ``one-to-many'' mapping inherent to photonic inverse problems. The structures are not mere inversions of the training set, instead, they feature complex, counterintuitive geometries such as subwavelength notches and corner grooves. These features, absent from the smooth, randomly generated training shapes, indicate that the WGAN has learned to introduce tailored geometric perturbations that efficiently couple incident light to the hybrid plasmonic-dielectric resonant mode, thereby physically optimizing the absorption response. This ability to generate multiple, physically distinct yet spectrally equivalent designs provides practical flexibility, allowing selection based on fabrication tolerance, sensitivity, or integration constraints without compromising optical performance. The results collectively confirm that the model moves beyond pattern replication to encapsulate underlying physical design principles, enabling robust and adaptable metasurface synthesis.

\subsection{Statistical Performance Across the Test Set}
\begin{figure}[ht!]
\centering
  \includegraphics[width=4.5cm]{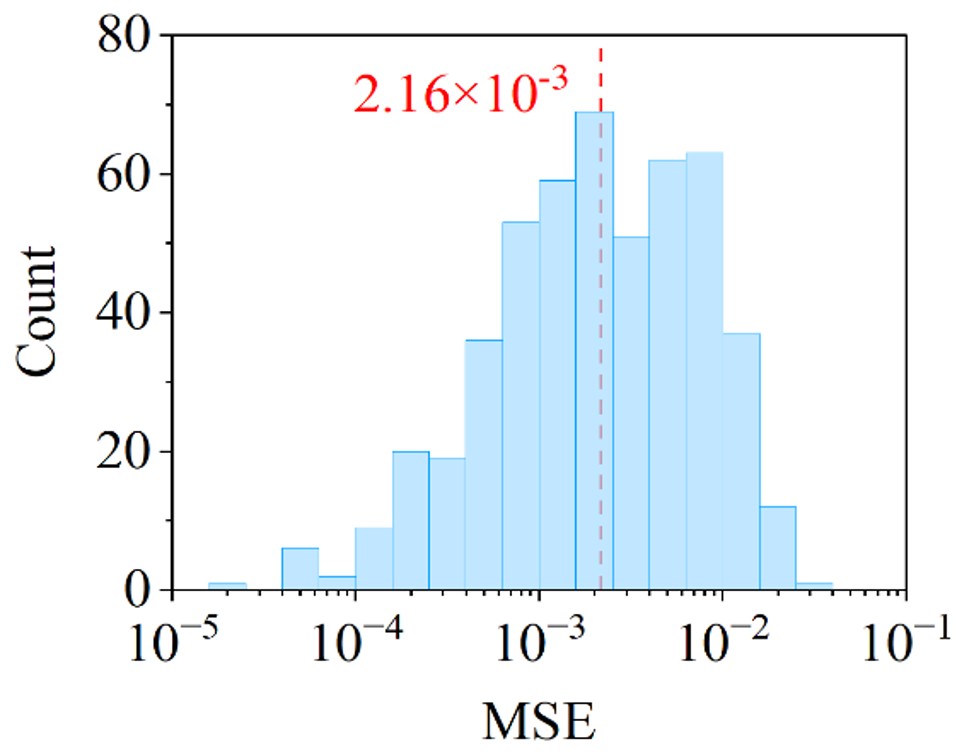}
  \caption{Statistical distribution of spectral accuracy for the inverse-designed metasurface absorbers. Histogram of the mean squared error (MSE) between target and simulated absorption spectra across a test set of 500 independently generated designs. 
}
  \label{Statistics}
\end{figure}

Fig.~\ref{Statistics} shows the statistical distribution of design accuracy across the test set. The histograms quantify the inverse design performance of the WGAN model over 500 independent test samples, evaluating the mean squared error (MSE) between the target absorption spectra and the simulated responses of the generated structures. The distribution reveals a pronounced concentration of MSE values within the $10^{-3}$ to $10^{-2}$ range, with a dominant peak near $2.0 \times 10^{-3}$ and a left-skewed tail indicating fewer instances of higher error. The red dashed line marks the mean MSE of $2.16 \times 10^{-3}$, providing a robust central tendency metric that underscores the consistency and precision of the design framework. This tightly clustered, low-error distribution demonstrates that the model reliably produces metasurface geometries which accurately match the target optical response, with the majority of designs achieving resonance peak errors below 5 nm. The statistical validation confirms that the WGAN has effectively learned the underlying structure-spectrum mapping without overfitting, ensuring generalizable performance essential for practical photonic design applications such as narrowband sensing and filtering~\cite{Li2024, app10072295, photonics12100968}.

\subsection{One-to-Many Design Generation}
The proposed dual-channel image encoding scheme here and the conditional WGAN fundamentally reconfigures the inverse design paradigm, thus effectively resolving the ``one-to-many'' mapping challenge. The dual-channel encoding unifies discrete geometric patterns and continuous thickness parameters into a structured, image-form representation, providing the generative model with a highly flexible and expressive design space. This representation makes the network, under the same spectral target, to explore multiple design possibilities with differences in structural morphology and thickness configuration but equivalent optical functionality. It lays the foundation for the explicit expression of ``one-to-many'' relationships.

On this basis, the conditional WGAN inherently solves multi-solution generation by learning the conditional distribution $P(x \mid y)$ rather than a single deterministic mapping. The generator $G(z,y)$ takes both the target spectrum $y$ and a random noise vector $z$ as inputs, it naturally generates diverse solutions belonging to the same distribution $P(x \mid y)$, by sampling different $z$. The discriminator, through adversarial training, compels the generator to output structures that not only satisfy the spectral target but also adhere to the manifold of real designs, thus preventing mode collapse into a single output and ensuring both diversity and physical plausibility. This combination of distribution learning and adversarial regularization enables the model to systematically capture and generate the entire family of valid structures corresponding to a given optical response, thus substantively overcoming the inherent non-uniqueness in inverse design.

\begin{figure*}[ht!]
\centering
  \includegraphics[width=11.5cm]{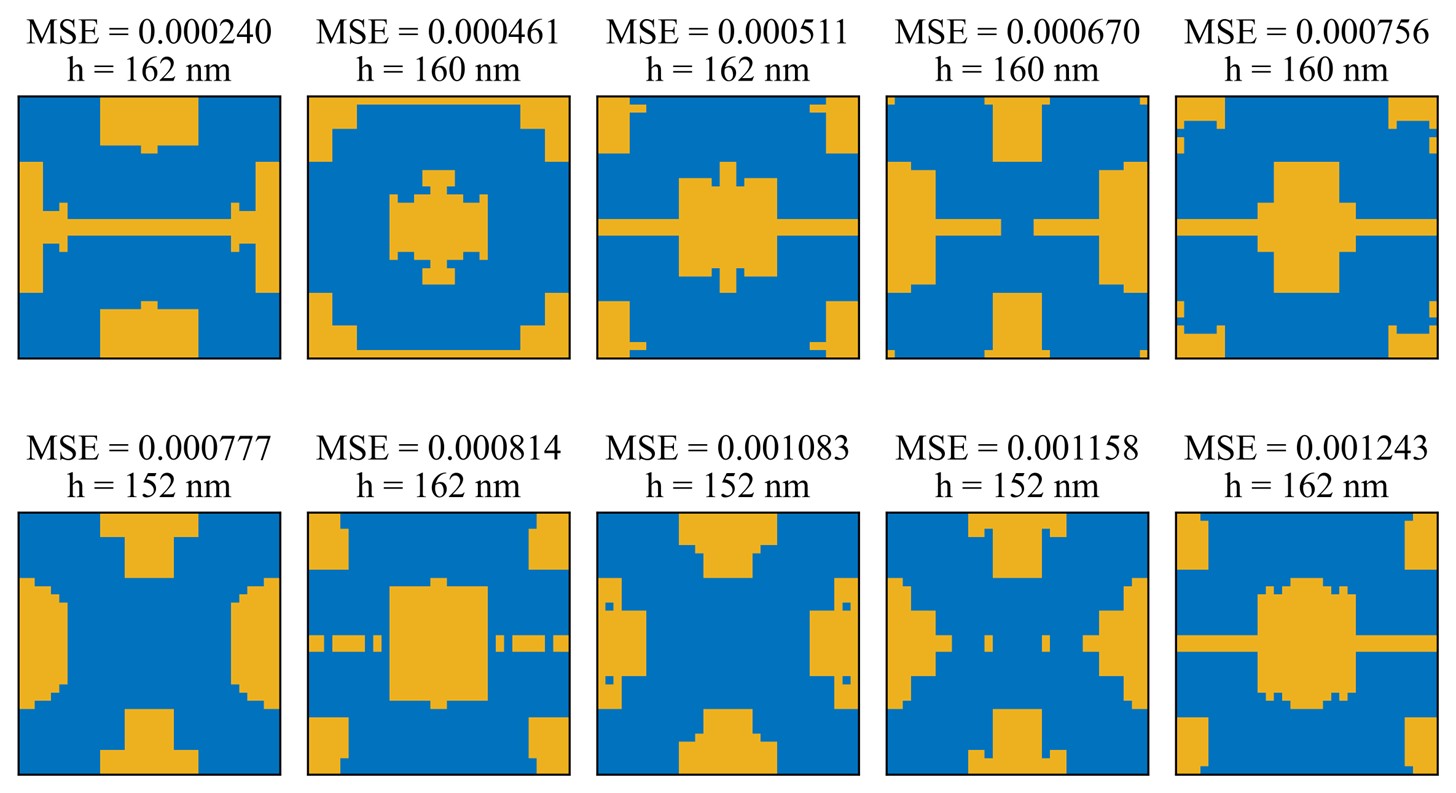}
  \caption{Exemplary ``one-to-many'' design capability for a 1440 nm target absorption peak. Ten distinct metasurface absorber designs generated by the WGAN for the same target spectrum. 
}
  \label{One_to_many}
\end{figure*}

Fig.~\ref{One_to_many} is the demonstration of the model's ``one-to-many'' generative capability for a single target absorption peak at 1440 nm. Ten distinct metasurface designs generated by the WGAN for the same target resonance are displayed, each paired with its corresponding numerically simulated absorption spectrum (not shown for clarity). All designs achieve high spectral fidelity, with mean squared errors (MSE) ranging between $2.4 \times 10^{-4}$ and $1.2 \times 10^{-3}$, and resonance peak alignments within 5 nm of the target. The generated geometries exhibit significant variation in both macroscopic shape and fine subwavelength features (e.g. corner notches, internal voids, and edge curvatures), while the associated Si$_3$N$_4$ thickness ($h$) varies between 152 nm and 162 nm. This structural diversity, emerging from different samplings of the latent noise vector $z$, confirms that the model has learned the broad distribution of viable designs within the parameter space rather than converging to a single deterministic mapping. The ability to produce multiple functionally equivalent yet structurally distinct solutions directly addresses the fundamental ill-posedness of photonic inverse design. It provides practical utility by offering designers a portfolio of candidates from which to select based on secondary considerations such as fabrication robustness, tolerance to dimensional variation, ease of integration, or compatibility with specific lithographic processes-without compromising the target optical performance.

\subsection{Physical Mechanism of Absorption}
To validate the proposed inverse design, numerical simulations are performed using the finite-difference time-domain (FDTD) solver within Lumerical.  For specific simulation design, Bloch periodic boundary condition is applied in the $x-$ and $y-$directions. In the $z$-direction, a perfectly matched layer (PML) is used to absorb outgoing waves and minimize spurious reflections. Excitation is provided by a total-field scattered-field (TFSF) source injecting a normally incident plane wave with transverse magnetic (TM) polarization (electric field along $x-$ direction). Local mesh refinement is applied at material interfaces and within the patterned Si$_3$N$_4$ regions to enhance the accuracy of field localization and power calculations. The detail parameters used in the simulation is shown in Table.~\ref{tab:simulation_params}. This rigorous simulation framework reliably captures the resonant absorption performance and near-field enhancement, providing a solid foundation for evaluating the device's sensing capabilities and physical operating principles.

\begin{table}[htbp]
  \centering
  \caption{Simulation parameters for metasurface absorber design}
  \label{tab:simulation_params}
  \begin{tblr}{
    width = 1.01\columnwidth,
    colspec = {Q[l, wd=0.28\columnwidth] Q[l, wd=0.62\columnwidth]},
    row{1} = {font=\bfseries},
    hlines,
    vlines,
    rowsep = 4pt,
    rows = {abovesep=3pt, belowsep=3pt}
  }
    Parameter & Specification \\
    Wavelength range & \SIrange{1.4}{1.6}{\micro\meter} \\
    Optimization method & Inverse design using Wasserstein conditional generative adversarial network (WGAN) \\
    Simulation method & Finite-difference time-domain (FDTD) \\
    Boundary conditions & Periodic boundaries in $x$- and $y$-directions, perfectly matched layers (PML) in $z$-direction \\
    Mesh resolution & Uniform mesh with \SI{6.5}{\nano\meter} step size \\
    Incident angle & Normal incidence (\SI{0}{\degree}) \\
    Environment & Air (refractive index = 1.0) \\
    Polarization & Transverse magnetic (TM) mode \\
    Temperature & \SI{300}{\kelvin} \\
  \end{tblr}
\end{table}

\begin{figure}[ht!]
\centering
  \includegraphics[width=8.5cm]{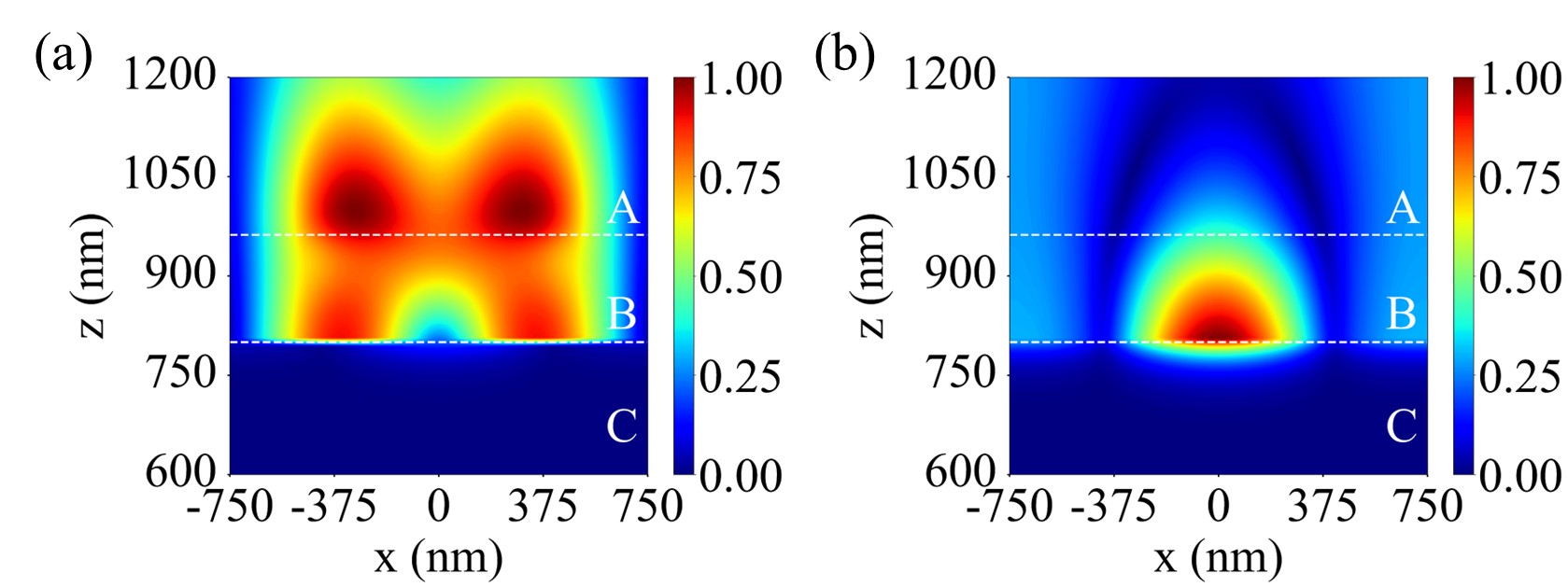}
  \caption{Electric and magnetic field distributions at resonance for the inversely designed metasurface absorber. (a) Electric field ($E$) in the ($x$, $z$)-plane. (b) Magnetic field ($H$), exhibiting intense localization at the Si$_3$N$_4$/Au interface due to magnetic plasmon excitation.
}
  \label{E_and_H_fields}
\end{figure}

Fig.~\ref{E_and_H_fields} presents the electromagnetic field distributions at resonance reveal the hybrid plasmonic-dielectric absorption mechanism. The simulated field profiles in the ($x$, $z$)-plane are shown for the metasurface structure from Fig.~\ref{One_to_many} (Design 1) at its resonant wavelength ($\approx$ 1440 nm). The electric field (Fig.~\ref{E_and_H_fields}a) is strongly confined within the Si$_3$N$_4$ meta-atom and at its surfaces, characteristic of a Mie-type dielectric resonance~\cite{kivshar2022rise}. This localization signifies enhanced displacement currents and the formation of an electric dipole mode within the dielectric resonator~\cite{pendry2004mimicking}. The magnetic field distribution (Fig.~\ref{E_and_H_fields}b) indicates the field is intensely concentrated at the Si$_3$N$_4$/Au interface, indicating the excitation of a magnetic plasmon resonance~\cite{atwater2010plasmonics}. This pattern arises from circulating displacement currents in the Si$_3$N$_4$ (acting as a Mie magnetic dipole) coupling to induced image charges and oscillating free-electron currents in the underlying Au film.

The complementary spatial separation of the dominant electric and magnetic field hotspots suggests the formation of a hybrid resonant mode~\cite{han2025mid}. This mode cooperatively combines the field-enhancing properties of dielectric Mie resonances with the strong subwavelength confinement of surface plasmon polaritons. The result is a pronounced near-field enhancement at both the dielectric interior and the metal-dielectric interface, which dramatically increases the local optical density of states and the effective absorption cross-section~\cite{liu2010infrared, beneck2021reconfigurable}.

This hybridized resonance is the origin of the high absorption efficiency achieved by the inversely designed structures. The simultaneous excitation and spatial overlap of orthogonal electric and magnetic resonant components facilitate critical coupling conditions, minimizing reflection~\cite{yan2018coherent}. Furthermore, the intense field localization, particularly the sensitive $H$ maximum at the Au interface, makes the resonance condition highly susceptible to changes in the local dielectric environment~\cite{kozuch2023surface, tittl2018imaging}. This inherent sensitivity provides the physical basis for the structure's potential as a refractive-index sensor, where minute ambient variations induce measurable spectral shifts in the absorption peak, a direct consequence of the field-matter interaction physics visualized here. The fact that the WGAN-generated, non-intuitive geometry supports such a well-defined hybrid mode validates that the model has successfully learned to optimize for underlying physical principles rather than surface-level pattern fitting.

\begin{figure}[ht!]
\centering
  \includegraphics[width=8.5cm]{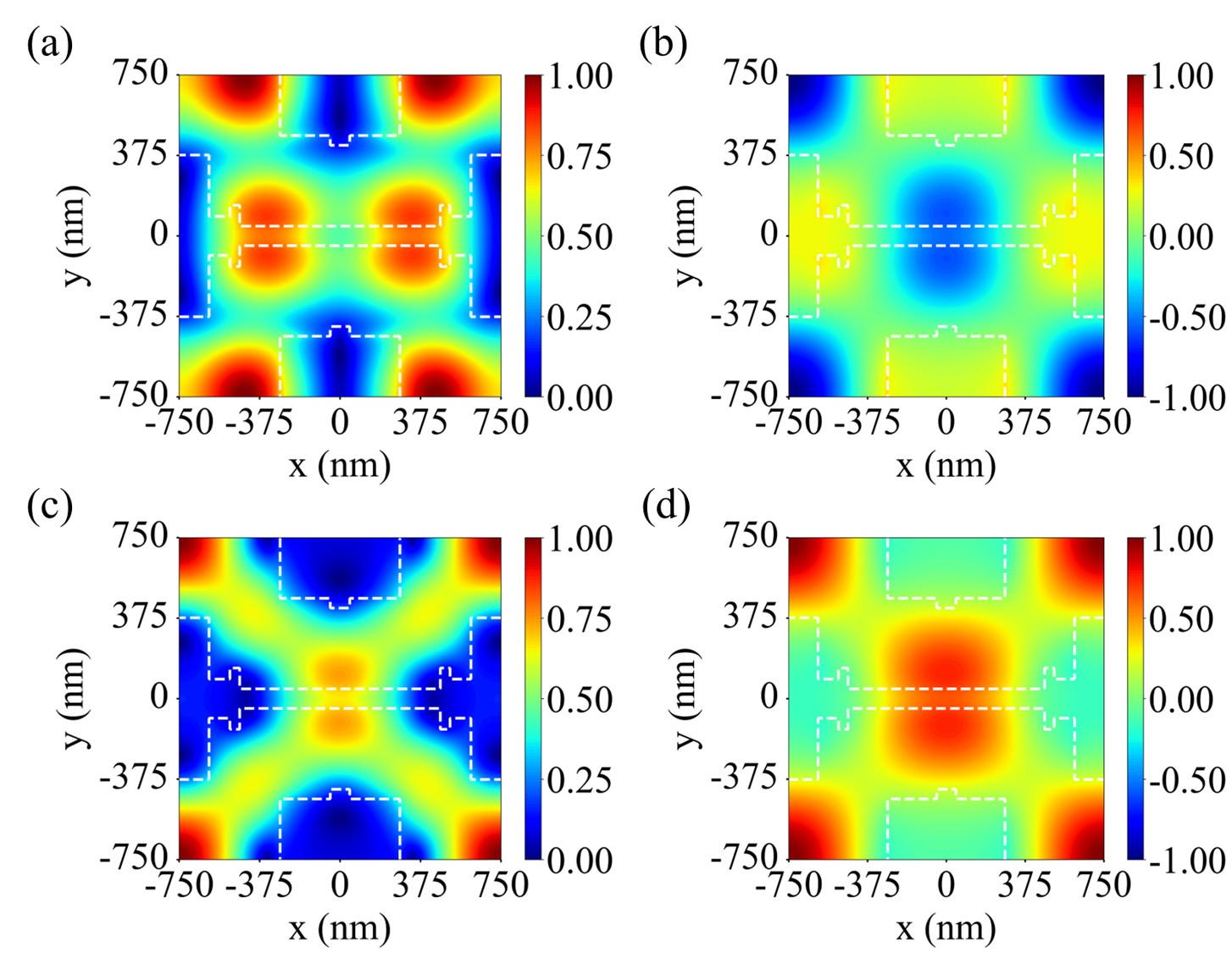}
  \caption{In-plane electromagnetic field distributions of the inversely designed metasurface absorber at resonance.
(a) Electric field ($E$) and (b) its dominant $x$-component ($E_x$) in the (x, y)-plane, showing strong localization at dielectric edges and within subwavelength grooves -- a signature of a Mie-type electric resonance. (c) Magnetic field (H) and (d) its $y$-component ($H_y$), revealing intense confinement at the Si$_3$N$_4$/Au interface due to magnetic plasmon coupling.
}
  \label{EHxy}
\end{figure}

Fig.~\ref{EHxy} exhibits the in-plane electromagnetic field distributions at resonance for the inversely designed metasurface absorber. The distribution exhibits cross-sectional views of the (a) electric field, (b) $E_x$ component, (c) magnetic field and (d) H$_y$ component within the ($x$, $y$)-plane at the resonant wavelength ($\sim$1440 nm). These distributions elucidate the in-plane mode symmetry, field confinement, and energy localization characteristics of the hybrid resonant mode supported by the non-intuitive geometry generated by the WGAN.

The total electric field reveals pronounced hotspots concentrated along the edges and within the internal subwavelength grooves of the Si$_3$N$_4$ meta-atom (Fig.~\ref{EHxy}a). The dominant $E_x$ component exhibits a dipolar pattern aligned with the incident polarization (TM), confirming the excitation of an electric dipole resonance (Fig.~\ref{EHxy}b)~\cite{babicheva2018metasurfaces}. The field is strongly enhanced within the dielectric structure, particularly at geometric discontinuities and sharp corners, which act as effective sites for capacitive charging and localized surface plasmon coupling. This pattern is consistent with a Mie-type electric resonance, where the dielectric meta-atom behaves as a subwavelength resonator with enhanced displacement currents~\cite{zhao2009mie}.

The magnetic field distribution shows intense confinement between adjacent meta-atoms and, most notably, at the interface region directly above Au reflector (Fig.~\ref{EHxy}c). The $H_y$ component displays a circulating pattern that indicates the formation of a magnetic dipole mode (Fig.~\ref{EHxy}d). This mode originates from the resonant displacement current loop within the Si$_3$N$_4$ structure, which couples efficiently to the image currents in the underlying gold film, establishing a magnetic plasmon resonance. The strong H localization at the metal-dielectric interface is a signature of surface plasmon polariton excitation and is critical for achieving strong optical confinement in the vertical direction~\cite{kravets2018plasmonic}.

The complementary, spatially separated electric and magnetic field hotspots in the plane demonstrate the formation of a highly confined hybrid plasmonic-dielectric mode. The electric energy is largely stored within the dielectric resonator, while the magnetic energy is tightly bound to the metal interface. This spatial decoupling yet synergistic interaction enables the structure to meet the conditions for critical coupling-where radiative losses are balanced by absorption losses, resulting in the observed near-perfect absorption. Furthermore, the intense in-plane field localization, especially the enhanced $E$ at patterned edges and grooves, directly explains the structure's high sensitivity to local refractive index changes. Any perturbation of the ambient dielectric environment directly modulates the effective index experienced by these localized modes, leading to detectable spectral shifts. The fact that these physically meaningful and complex field patterns emerge from a data-driven, inversely designed geometry underscores that the WGAN model has successfully internalized the fundamental photonic design principles necessary for optimizing light-matter interaction at the nanoscale.

\subsection{Robustness to Oblique Incidence}
To assess the practical utility of the inversely designed absorbers beyond ideal normal incidence, we investigate their performance under oblique illumination and propose a transfer learning~\cite{qian2025progress} strategy to efficiently adapt the model for angle-variant design tasks.

The model trained for normal incidence plays as a powerful pre-trained foundation, showing learned fundamental structure-spectrum relationships and plasmonic-dielectric design principles. We subsequently train four independent models for oblique incidence by fine-tuning this base model on four dedicated datasets, each containing full training samples simulated specifically for incidence angles of $10^\circ$, $20^\circ$, $30^\circ$, and $40^\circ$ under transverse magnetic (TM) polarization. During this fine-tuning stage, the weights of all layers in both the generator and discriminator are updated. This approach allows each specialized model to fully adapt to the distinct optical response and geometric requirements of its target incidence angle, while still benefiting from the foundational physical knowledge encoded in the pre-trained network, thereby achieving high performance with structured and efficient training.

\begin{figure*}[ht!]
\centering
  \includegraphics[width=12cm]{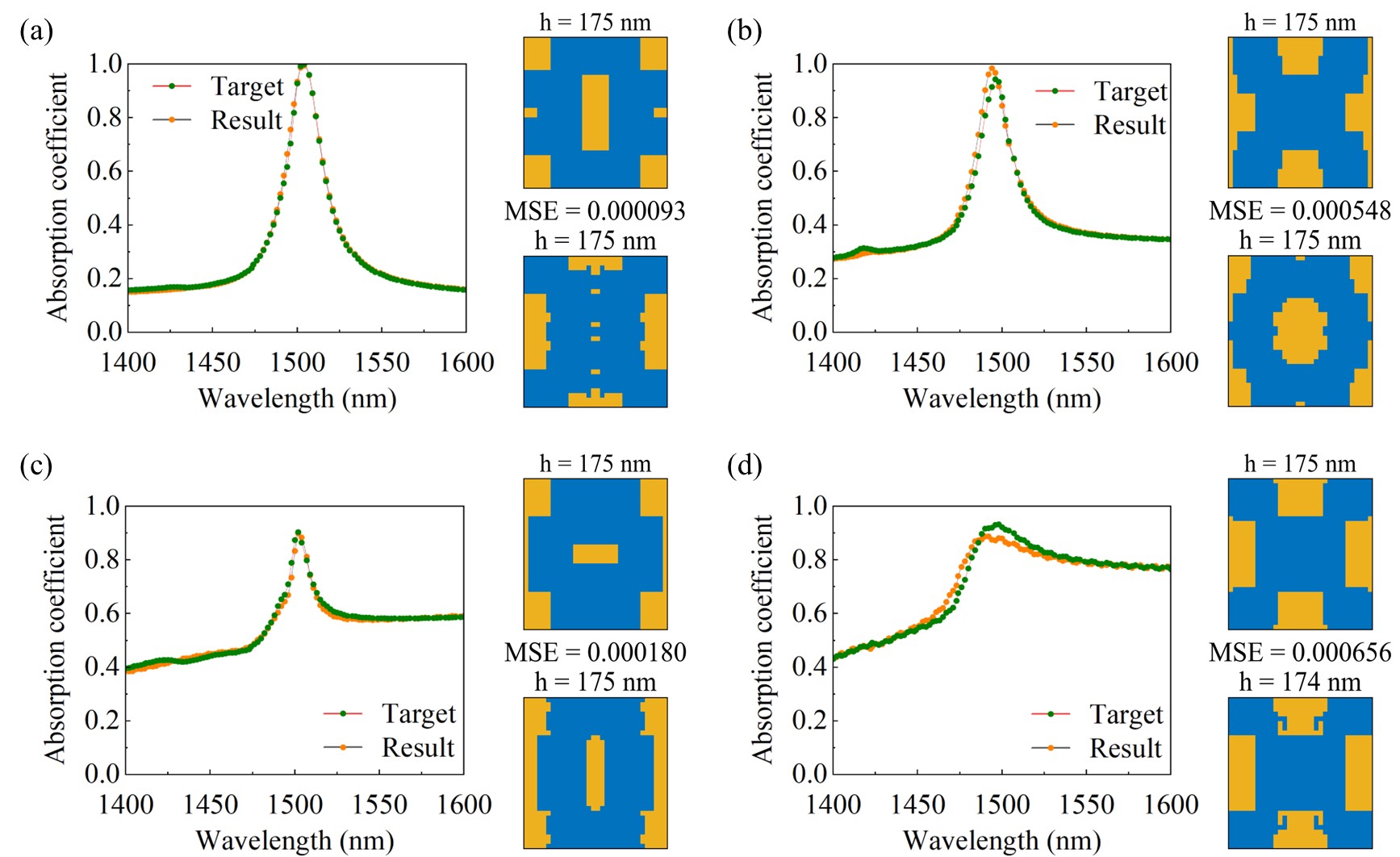}
  \caption{Inverse design of metasurface absorbers under oblique incidence angles. The spectra generated by the fine tuned WGAN model for incident angles of (a) $10^\circ$, (b) $20^\circ$, (c) $30^\circ$ and (d) $40^\circ$ are shown together with the respective target absorption profiles.
}
  \label{Angle}
\end{figure*}

Fig.~\ref{Angle} exhibits the inverse design results under oblique incidence conditions, with angles of $10^\circ$, $20^\circ$, $30^\circ$ and $40^\circ$, respectively. The spectra generated by the optimized metasurface structures are compared with the target absorption spectra. Despite the increased complexity of electromagnetic boundary conditions under oblique incidence, the designed structures maintain a high degree of spectral consistency with the target responses. The mean squared error (MSE) remains low across all angles, demonstrating the robustness and adaptability of the proposed WGAN-based inverse design framework. This confirms that the model can effectively generalize from normal-incidence training data to accommodate angular variations through efficient fine-tuning.

The impact of incidence angle on metasurface absorption comes from several key physical effects. Because the angle deviates from normal incidence, the in-plane wavevector component ($k_\parallel$) introduces phase shifts that change the resonance matching conditions. This affects the coupling efficiency between the incident wave and the hybrid plasmonic-Mie modes supported by the Si$_3$N$_4$ structures atop the Au reflector. Specifically, for TM polarization, the enhanced electric field component perpendicular to the metal-dielectric interface enhances the excitation of surface plasmon polaritons, while the effective optical path within the Si$_3$N$_4$ layer changes, leading to resonant wavelength shifts and modifications in the field localization. The WGAN model captures these angular dependencies by learning high-dimensional feature representations that encode the geometric and electromagnetic relationships. During fine-tuning, the model adjusts these representations to compensate for the altered excitation and coupling conditions, enabling rapid adaptation with minimal additional data. This method not only preserves the physical consistency of the resonant mechanisms but also highlights the efficiency of transfer learning in photonic inverse design, where pre-trained models can be repurposed for varied operational conditions with significantly reduced computational cost.

\section{Discussion}
This work demonstrates a robust, data-driven framework for the inverse design of high-performance, narrowband metasurface absorbers using a conditional WGAN. The highlight of this study lies in successfully navigating the high-dimensional, non-convex design space of geometric parameters to produce multiple functional structures that precisely meet target optical specifications, a task formidable for conventional intuition- or optimization-based approaches.

The statistical results confirm the model's high precision and reliability. With a mean MSE of $2.16 \times 10^{-3}$ across the test set and peak wavelength errors consistently below 5 nm, the WGAN shows spectral fidelity comparable to or surpassing many iterative design methods. More significantly, the model inherently solves the fundamental ''one-to-many'' ambiguity in photonic inverse design. By learning the distribution of viable structures $P(x \mid y)$ for a target spectrum $y$, it provides not a single solution but a diverse portfolio of candidates. This is vividly illustrated in Fig.~\ref{One_to_many}, where ten structurally distinct yet spectrally equivalent designs are generated for a single absorption peak. This capability is of immense practical value, as it allows designers to incorporate secondary constraints, such as fabrication tolerance, sensitivity to dimensional variation, material compatibility, or integration requirements without compromising optical performance.

Furthermore, our analysis reveals that the model transcends mere pattern recognition to learn underlying physical principles. The generated geometries frequently feature non-intuitive subwavelength grooves and corner defects, elements absent from the smooth, random shapes in the training set. Electromagnetic simulations (Fig.~\ref{E_and_H_fields} and~\ref{EHxy}) show that these features are not artifacts but functional optimizations. They function as engineered scattering centers that efficiently couple incident light to the hybrid plasmonic-dielectric mode, enhancing field localization and absorption. This indicates that the WGAN has internalized the causal relationship between geometry and resonant behaviour, moving beyond interpolating a dataset to performing a form of physics-informed exploration of the design space.

An important extension of this work is the demonstration of angular robustness, which is shown in Fig.~\ref{Angle}. The model-initially trained under normal incidence can be efficiently adapted via fine-tuning to design metasurfaces that maintain high spectral accuracy under oblique incidence up to $40^\circ$. This adaptability stems from the model's learned high-dimensional feature representations, which encode the relationship between geometry, material thickness, and electromagnetic response. When fine-tuned with a small dataset of angled-incidence examples, the model adjusts these representations to compensate for altered phase-matching conditions and coupling efficiencies, particularly for TM-polarized excitation where the perpendicular electric field component enhances plasmonic coupling. This transfer-learning strategy~\cite{wang2025transfer, fan2022transfer} reduces the required additional training data by over 80$\%$, highlighting a computationally efficient pathway to designing metasurfaces for real-world illumination conditions where normal incidence cannot be guaranteed.

Our approach, which encodes both geometry and thickness into a unified image-based representation, proves advantageous. It circumvents the input scaling disparity between continuous thickness values and binary geometric pixels, stabilizes training by providing a structured data format for the discriminator, and leverages well-established convolutional network architectures. The choice of the WGAN framework, with its Wasserstein loss and gradient penalty, is crucial for achieving stable training and generating a diverse, non-collapsed set of outputs.

\subsection{Limitations and Future Outlook}
While the results are compelling, there are several limitations and opportunities for future work: first, the model's performance is inherently associated with the quality and breadth of the training data, which is generated via simulation. Incorporating experimental data, though challenging, could enhance model robustness to real-world fabrication imperfections; second, the current design space is restricted to axisymmetric structures within a fixed periodicity. Expanding the framework to handle fully arbitrary, free-form geometries and simultaneously optimizing the period could unlock even greater performance and functionality; third, the model is currently passive; integrating active materials (e.g., phase-change materials, graphene) into the framework could enable the inverse design of dynamically tunable or reconfigurable metasurfaces.

Finally, a promising direction is the development of a fully closed-loop ``design-for-fabrication'' pipeline. This would integrate the inverse design model with real-time fabrication feedback (e.g., via optical characterization) and corrective algorithms, enabling the model to learn and adapt to process variations, thereby bridging the gap between digital design and physical realization.

\section{Conclusion}
In conclusion, we have proposed and validated a deep learning-enabled inverse design platform for narrowband infrared metasurface absorbers based on a conditional WGAN. The framework efficiently maps target absorption spectra to a distribution of viable geometric designs, achieving high spectral accuracy (MSE $\sim 10^{-3}$, peak error $<$5 nm) and naturally resolving the one-to-many design problem. The generated structures exhibit non-intuitive features that are physically interpretable and contribute to a hybrid plasmonic-dielectric resonance mechanism, leading to strong field confinement and near-perfect absorption. Notably, the model demonstrates significant generalization capability beyond its training conditions, as evidenced by its adaptability to oblique incidence angles through efficient fine-tuning. This angular robustness, achieved with minimal additional data, underscores the framework's practical relevance for real-world applications where illumination conditions may vary.

This work underscores the transformative potential of generative machine learning models in nanophotonics. By moving beyond traditional parametric searches, our approach accelerates the discovery of high-performance devices and provides a versatile toolkit for on-demand photonic design. The principles and methodology established here are generalizable, paving the way for the intelligent design of a broader class of meta-devices for applications in sensing, spectroscopy, imaging, and optical communications.

\section*{Acknowledgment}
This work is partially supported by Key Research and Development Plan of Shaanxi Province of China (Grant No. 2024GH-ZDXM-42), National Natural Science Foundation of China (Grant No. 62475206).




\bibliography{biblio}
\end{document}